\documentclass{article}
\usepackage[utf8]{inputenc}
\usepackage{authblk}
\usepackage{setspace}
\usepackage[margin=1.25in]{geometry}
\usepackage{graphicx}
\graphicspath{{./figures/}}
\usepackage{subcaption}
\usepackage{amsmath}
\usepackage[mathlines]{lineno}
\usepackage{siunitx}
\usepackage[normalem]{ulem}

\usepackage[
  style=nejm,
  citestyle=numeric-comp,
  sorting=none
]{biblatex}
\title{Galvanic Vestibular Stimulation in Latent Space}

\author[1*]{Zhi Liu}
\author[2,3,4]{Tatsuki Fushimi}
\author[2,3,4,5]{Yoichi Ochiai}

\affil[1]{Graduate School of Comprehensive Human Sciences, University of Tsukuba, Tsukuba, 305-8550, Japan}
\affil[2]{Institute of Library, Information and Media Science, University of Tsukuba, Tsukuba, 305-8550, Ibaraki, Japan}
\affil[3]{R\&D Center for Digital Nature, University of Tsukuba, Tsukuba, 305-8550, Ibaraki, Japan}
\affil[4]{Tsukuba Institute for Advanced Research (TIAR), University of Tsukuba, 1-1-1 Tennodai, Tsukuba, Ibaraki, 305-8577, Japan}
\affil[5]{Pixie Dust Technologies, Inc., Chuo-ku, 104-0028, Tokyo, Japan}
\affil[*]{Address correspondence to: shawnliu@digitalnature.slis.tsukuba.ac.jp}

\date{}

\begin{document}

\maketitle

\begin{abstract}
Galvanic vestibular stimulation (GVS) is widely used to modulate self-orientation, balance, and motion perception; the discriminability of frequency-encoded cues further suggests its potential as a standalone modality for embodied feedback. However, synthesizing GVS waveforms congruent with target events or bodily states remains challenging: GVS waveforms combine current direction, intensity, duration, and onset/offset transitions, yet how these parameters jointly shape users' perceptual and associative responses remains underexplored. To address this gap, we contribute a dataset linking GVS waveforms to free-form experience descriptions and a retrieval-guided generative model for synthesizing candidate waveforms from target descriptions. The dataset comprises 100 GVS waveforms and 1,526 valid free-form sensation descriptions from 16 participants. Semantic analysis revealed diverse motion- and force-related sensations, localized bodily sensations, and situational associations. Compared with a participant-preserving permutation baseline, descriptions elicited by the same waveform covered fewer semantic categories (\num{8.18} vs.\ \num{9.45}) and exhibited a higher dominant-category proportion (\SI{26.97}{\percent} vs.\ \SI{21.25}{\percent}; both \(P < 0.001\)). Building on this dataset, we implemented the generative model as a retrieval-guided one-dimensional convolutional variational autoencoder. An independent behavioral study recruited 10 participants who had not contributed to dataset collection. Performance in discriminating congruent from incongruent waveform--visual cue pairings was significantly above chance (accuracy \(= \SI{63.33}{\percent}\), \(d' = 0.70\), and \(P < 0.001\)). Together, these findings demonstrate the feasibility of text-conditioned GVS synthesis and support the development of GVS as a programmable modality for semantically congruent embodied feedback across interactive scenarios.
\end{abstract}

\section*{Introduction}
Galvanic vestibular stimulation (GVS) is a non-invasive method in which weak currents are delivered via electrodes positioned near the mastoid processes, thereby altering activity in vestibular afferents arising from both the semicircular canals and otolith organs~\cite{Base2025,Kwan2019}.
Given GVS's ability to non-invasively influence balance perception, spatial orientation, and postural control, it has been used as a tool for probing human vestibular function and eliciting vestibular reflexes~\cite{GVSeffect,Keywan2018}. Prior work has quantified GVS-induced effects using perceptual~\cite{GVSgallagher2023quantifying} and postural measures~\cite{GVSwardman2003effects,LRtilt,Demura2006}, including perceived rotation and verticality, body sway and direction-dependent displacement.
The ability to evoke and quantify these responses has also motivated research on GVS in medical and rehabilitation contexts~\cite{Dizzness,Parkinson,Fujimoto2016}.
In recent years, GVS has also attracted attention in entertainment contexts. Prior work has explored its use in device-affordance-based digital games~\cite{balanceNinja}, perception enhancement~\cite{rollerCoaster,GVSbike}, and perception manipulation~\cite{RedirectionW,RedirectHand,Aoyama2015}.
These studies suggest that GVS can serve as an embodied feedback channel by directly modulating users' vestibular and bodily sensations, thereby enabling perceptual effects that are difficult to convey through conventional modalities, such as visual, auditory, and tactile.
Beyond conveying information related to self-orientation, GVS can also function as an alternative display modality. Smith et al.~\cite{GVSfeedback} showed that users could reliably discriminate frequency-encoded GVS cues across standing, walking, passive-motion, and noisy conditions.
This finding suggests that GVS can also leverage other distinguishable sensations induced by electrical stimulation as interaction cues.
As a variant of tDCS~\cite{tdcs}, current transitions also introduce stimulation-associated sensations that are relevant to the applied waveform. Rapid changes in stimulation current can produce transient cutaneous sensations at the electrode sites, such as burning and tingling~\cite{GVSgoel2015,GVSutz2011minor}.
Therefore, in conventional orientation and balance related tasks, stimulation is generally ramped up and ramped down at the beginning and end of the session to prevent cutaneous discomfort~\cite{FrontiersGVSreview}. 
The stimulation waveform involves multiple parameters, including current direction, intensity, onset/offset transition profile, and duration. However, onset and offset ramps are often specified as fixed implementation settings~\cite{ramp2006,ramp2018,ramp2019}, quantitative evidence on how variations in ramp configuration shape the sensations elicited by GVS remains limited.
Zhao et al.~\cite{xiaoyu} employed a graphical user interface (GUI) for customizing GVS waveforms, enabling participants to adjust waveform parameters for different use scenarios and explore the bodily sensations elicited by different waveforms. However, systematic evidence linking GVS waveform parameters to their elicited sensations remains limited. To the best of our knowledge, no publicly available dataset pairs GVS waveforms with free-form textual descriptions of sensations.
Such a dataset would enable analysis of the subjective experiences associated with different waveform configurations, the semantic distinguishability of these experiences, and their consistency across users.

Moreover, recent advances in machine learning make paired waveform--description data suitable for training conditional generative models that support data-driven candidate waveform design for a target verbal sensation description, thereby reducing reliance on manual tuning and repeated user testing in entertainment and other interactive applications.

From a modeling perspective, we formulate the task as natural-language-conditioned candidate GVS waveform generation: given a verbal description of a desired sensation, the goal is to generate waveforms that are associated with semantically similar experiences within an empirically observed stimulation space. This formulation does not assume a deterministic one-to-one mapping between a sensation description and a GVS waveform. Instead, it aims to model population-level waveform--sensation associations observed under a bounded experimental protocol.

Generative models have enabled time-series synthesis under different forms of conditioning information. For example, TTS-CGAN generates class-specific biosignal time series conditioned on predefined categorical labels~\cite{ttscgan}, whereas Time Weaver incorporates heterogeneous metadata, including categorical, continuous, and time-varying variables, to guide generation~\cite{TimeWaver}. More recent text-to-time-series methods use free-form natural-language descriptions as flexible generation conditions~\cite{VerbalTS}. However, directly applying end-to-end text-conditioned generation methods to GVS waveform generation remains challenging. Constructing paired waveform--description datasets is costly, and GVS responses may vary across users and stimulation conditions, making waveform--sensation associations noisy and potentially many-to-many~\cite{Base2025,Allred2024}. Under such limited-data conditions, generative models may overfit to observed samples or produce inadequate output diversity, and unlike visual content generation, erroneous outputs are difficult to identify through direct inspection~\cite{fewshot}.

Several approaches have been proposed to address data scarcity in time-series generation. 
Autoencoders (AEs) compress input data into latent representations and reconstruct the original data through a decoder~\cite{AEbase}. While variational autoencoders (VAEs) extend this framework by imposing a probabilistic structure on the latent space, enabling new samples to be generated by sampling and decoding latent variables~\cite{kingma2014auto}.
Zhang et al.~\cite{Zhang2024VAEAugmentation} combined VAE with metric learning for multivariate time-series augmentation. Their approach structures the latent space by encouraging informative relationships among samples and improves the fidelity, stability, and generalization of generated data under limited data availability. Another relevant direction is retrieval-augmented generation (RAG), which combines parametric generation with retrieved non-parametric memory~\cite{RAGDefine}. Retrieved examples can provide local empirical references that complement a generative model, particularly when training data are limited. For example, Wang et al.~\cite{SEDiff} used retrieved clinical context to guide text-conditioned ECG generation, improving waveform fidelity and text--waveform alignment. Similarly, Liu et al.~\cite{retrivl} retrieved relevant observed sequences and incorporated them into diffusion-based forecasting, improving prediction reliability in challenging cases. Unlike these prior settings, GVS waveform generation requires linking free-form descriptions of subjective sensations to stimulation waveforms while accounting for limited data and substantial inter-participant variability. Motivated by these findings, we combine a VAE-based temporal representation with semantic retrieval: the VAE models structured variation in waveform trajectories, whereas retrieval provides semantically relevant empirical references that help constrain generation under limited data.

This work makes three contributions. First, we construct a paired dataset comprising GVS waveforms and participants' free-form descriptions of the sensations elicited by those waveforms. Second, using this dataset, we examine whether descriptions associated with the same waveform are semantically consistent across participants, thereby characterizing the reproducibility and variability of waveform--sensation associations across users. Third, we develop a retrieval-guided 1D-CNN VAE framework for natural-language-conditioned candidate GVS waveform generation. The 1D-CNN VAE learns latent representations of temporal waveform structure, while semantic retrieval identifies previously observed waveforms associated with descriptions similar to a target sensation. The retrieved examples are weighted and fused to construct a generation reference that incorporates both semantic relevance and waveform characteristics. We further validate the applicability of the proposed framework with unseen users, allowing us to evaluate the accuracy and usability of the generated waveforms in a realistic setting.

\section*{Results}

\begin{table}[!b]
    \centering
    \small
    \caption{Summary of perceptual semantic categories and VAE latent-space evaluation metrics.}
    \label{tab:summary_results}

    \textbf{A. Semantic categories of perceptual descriptions evoked by GVS}

    \vspace{0.3em}

    \begin{tabular}{c p{9.2cm} c}
        \hline
        \textbf{Category} &
        \textbf{Representative keywords and phrases} &
        \textbf{No. of descriptions, \(n\)} \\
        \hline
        1  & unnoticeable; weak; hard to describe; no clear sensation & 177 \\
        2  & centrifugal; bicycle; rotating & 163 \\
        3  & single; sides; tug & 99 \\
        4  & gentle push; gentle touch; gentle tap & 62 \\
        5  & wind; gentle breeze; natural swaying motion & 48 \\
        6  & hit; teaser; colliding & 144 \\
        7  & pushed; forcefully; continuously & 69 \\
        8  & swing; sea; swaying boat & 104 \\
        9  & dizziness; dizzy; unsteady & 62 \\
        10 & veins; twitch; tremor & 118 \\
        11 & gentle hint; notification; incoming & 77 \\
        12 & bug; insect; bite & 112 \\
        13 & sudden; braking; premonition & 142 \\
        14 & dragged; pulled; suckers & 67 \\
        15 & head; turned; scalp & 82 \\
        \hline
        \textbf{Total} & & \textbf{1,526} \\
        \hline
    \end{tabular}

    \vspace{1em}

    \textbf{B. VAE latent-space evaluation metrics}

    \vspace{0.3em}

    \begin{tabular}{lcc}
        \hline
        \textbf{Evaluation criterion} &
        \textbf{Summary statistic} &
        \textbf{Value} \\
        \hline
        Local latent sensitivity &
        Median (90th percentile) &
        0.759 (1.183) \\

        Local perturbation waveform RMS &
        Median (90th percentile) &
        0.032 (0.051) \\

        Interpolation jump ratio &
        Mean (90th percentile) &
        1.247 (1.445) \\

        Interpolation path efficiency &
        Mean &
        1.022 \\

        Latent--waveform distance consistency &
        Pearson's $r$ &
        0.945 \\
        \hline
    \end{tabular}
\end{table}

\begin{figure}[htbp]
    \centering
    \includegraphics[width=\textwidth]{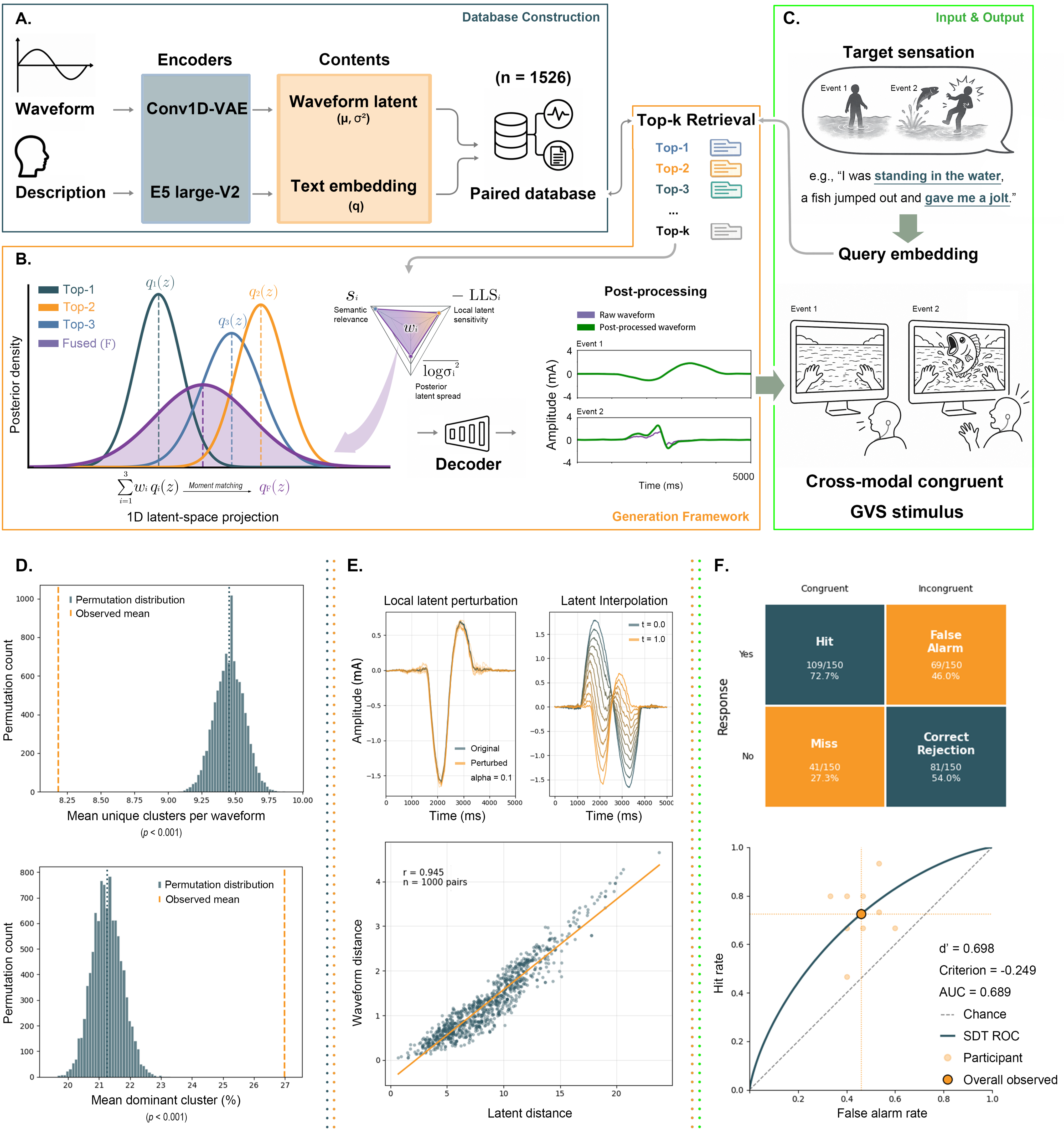}
    \caption{Overview of the cross-modal GVS generation and evaluation framework.
(A) Waveforms and textual descriptions are encoded into paired latent representations to construct a retrieval database.
(B) Top-$k$ latent distributions are fused, decoded, and post-processed to generate candidate GVS waveforms.
(C) A target sensation is converted into a query embedding and used to retrieve and generate a cross-modally congruent stimulus.
(D) Permutation analyses showed fewer unique semantic categories and a higher dominant-category proportion per waveform than expected by chance ($P < 0.001$), indicating waveform-dependent semantic-category associations.
(E) Latent-space perturbation, interpolation, and distance analyses evaluate waveform continuity and structural consistency.
(F) User-study outcomes are summarized by a stimulus--response matrix and an SDT ROC curve. The overall operating point (black-outlined orange marker) yielded $d' = 0.698$ and $c = -0.249$, with an AUC of 0.689.}
    \label{fig:vae_continuity}
\end{figure}

The proposed framework translates a free-form natural-language description of a desired sensation into a \(5\,\mathrm{s}\) GVS waveform. Each waveform comprises 250 discrete samples at a sampling rate of \(50\,\mathrm{Hz}\). The generated waveform is intended to evoke a sensation that is cross-modally congruent with the input description.
The framework first constructs a paired multi-modal database (Fig.~\ref{fig:vae_continuity}A). GVS waveforms and their perceptual descriptions are encoded into waveform and text representations, respectively. Given a target description, the framework retrieves semantically relevant waveform candidates. It then fuses their latent distributions and decodes the fused representation into a stimulation waveform. The waveform is further post-processed before final generation (Fig.~\ref{fig:vae_continuity}B). The post-processed waveform is then delivered as the final GVS stimulus (Fig.~\ref{fig:vae_continuity}C).

The three evaluations correspond to the three main stages of the framework. The perceptual pre-test evaluates the paired database constructed in Fig.~\ref{fig:vae_continuity}A by examining whether the waveform library contains systematic perceptual structure (Fig.~\ref{fig:vae_continuity}D). The latent-space analyses evaluate the retrieval, fusion, and decoding process in Fig.~\ref{fig:vae_continuity}B, by assessing local stability, interpolation continuity, and waveform-geometry preservation (Fig.~\ref{fig:vae_continuity}E). The independent behavioral experiment evaluates the final GVS stimuli shown in Fig.~\ref{fig:vae_continuity}C by testing their perceptual discriminability and cross-modal congruence with the intended sensations (Fig.~\ref{fig:vae_continuity}F).

\subsection*{Semantic category distribution and waveform-specific organization}
The waveform library used in this study consisted of 100 randomly generated GVS waveforms with varying temporal and amplitude characteristics (see Methods~\ref{sec:gui_validation} for details).
We first characterized how participant-generated descriptions were distributed across the selected semantic-category partition. We then examined waveform-specific semantic organization using the number of categories represented within each waveform, and the proportion of responses assigned to its dominant category. The waveform-level organization measures were compared against a participant-preserving permutation baseline.
The final analysis included 1,526 valid waveform-specific perceptions in text form (descriptions), from 16 participants, each of whom evaluated 100 GVS waveforms. On average, each waveform was associated with \(15.26 \pm 0.58\) valid descriptions.

Using the semantic categorization procedure, the participant-generated descriptions were summarized into 15 semantic categories. The resulting categories are summarized in Table~\ref{tab:summary_results}A. The category distribution showed a Shannon entropy of 3.82 bits, close to the theoretical maximum of 3.91 bits for 15 categories, corresponding to a normalized entropy of 0.98. Moreover, the largest category accounted for only 11.6\% of all descriptions. 

Next, we examined whether descriptions were organized by waveform identity. 
Within a given waveform, a smaller number of semantic categories indicates greater similarity in the sensations evoked by stimuli, whereas a higher proportion of the dominant semantic category indicates stronger convergence toward a common evoked sensation.

At the waveform level, semantic descriptions were distributed across an average of \(8.18\) semantic categories (\(SD = 1.59\)). The mean proportion of descriptions assigned to the dominant semantic category within each waveform was \(26.97\%\) (\(SD = 9.86\%\)).
Relative to a participant-preserving permutation null model based on 10,000 permutations, the actual descriptions showed fewer semantic categories per waveform than expected under permutation, \(M_{\mathrm{obs}} = 8.18\), \(M_{\mathrm{null}} = 9.45\), \(p_{\mathrm{perm,two\text{-}sided}} < 0.001\). The dominant semantic category also accounted for a larger proportion of responses than expected under the null model, \(M_{\mathrm{obs}} = 26.97\%\), \(M_{\mathrm{null}} = 21.25\%\), \(p_{\mathrm{perm,two\text{-}sided}} < 0.001\) (see Fig.~\ref{fig:vae_continuity}D). 

\subsection*{Latent-space continuity and geometric consistency}
The generation model used a 1D-CNN VAE to represent all GVS waveforms in a latent space. Because subsequent waveform generation operates within this space, we examined whether its geometry reflected smooth and organized variation in the decoded waveforms.
We assessed latent-space smoothness using three complementary analyses: local latent perturbation, latent interpolation, and latent--waveform distance correlation.

Median-case examples and the latent--waveform distance relationship, and it's corresponding quantitative metrics are summarized in 
Fig~\ref{fig:vae_continuity}E.
The local perturbation analysis showed that small changes in latent space generally produced limited changes in the decoded waveforms. Perturbed latent vectors decoded into waveforms that remained close to the waveform decoded from the original latent mean. As shown in Table~\ref{tab:summary_results}B, the median local latent sensitivity was 0.759, with a 90th percentile value of 1.183. The corresponding perturbation-induced waveform root--mean--square (RMS) had a median value of 0.032 and a 90th percentile value of 0.051, indicating that the absolute waveform-level changes caused by local latent perturbations were small.

Latent interpolation produced gradually changing decoded waveforms between the two endpoints.
In the representative interpolation path, the decoded waveform gradually changed as the interpolation coefficient increased from $t=0$ to $t=1$. Across sampled interpolation paths, the mean jump ratio was 1.247, with a 90th percentile value of 1.445, suggesting that adjacent waveform changes along the interpolation paths were relatively uniform and did not show strong discontinuities. The mean interpolation path efficiency was 1.022, close to the ideal value of 1, indicating that decoded interpolation trajectories were close to direct paths between endpoint waveforms.
The latent--waveform distance analysis further showed that the learned latent geometry was strongly aligned with waveform-level similarity. 
Pairwise latent distances were strongly correlated with decoded waveform distances, with a correlation coefficient of 0.945 (Fig.~\ref{fig:vae_continuity}E). 
This indicates that nearby latent vectors tended to decode into similar waveforms, whereas larger latent distances tended to correspond to larger waveform differences. 

\subsection*{Perceptual Evaluation of Model-generated GVS Stimuli}
To assess the practical validity of the generated waveforms, we conducted an independent behavioral evaluation with participants who had not contributed to the dataset. The evaluation tested whether they could distinguish congruent from incongruent pairings between narrative scenes and model-generated GVS waveforms (see Methods for details).

Through a Unity-based GUI, participants were asked to evaluate whether the perceptual stimuli evoked by each model-generated GVS waveform were consistent with the paired narrative scene.
Each participant completed 30 trials, consisting of 15 \textit{congruent pairs} and 15 \textit{incongruent pairs} randomly sampled from separate pools of 50 pairs per type, resulting in 300 trials in total.

Signal detection indices were calculated from participants' performance in the binary congruence-judgment task, including hit rate, false alarm rate, balanced accuracy, sensitivity ($d^{\prime}$) and response criterion ($c$). Statistical significance was assessed using one-sample tests against chance level and paired-samples comparisons.
Across all trials, participants made 190 correct responses out of 300, corresponding to an overall accuracy of 63.33\%. For \textit{congruent pairs}, the hit and miss rates were 72.67\% ($109/150$) and 27.33\% ($41/150$), respectively. For \textit{incongruent pairs}, the false alarm and correct rejection rates were 46.00\% ($69/150$) and 54.00\% ($81/150$), respectively. The resulting balanced 
accuracy was 63.33\%.

Signal detection analysis yielded a sensitivity index of $d^{\prime} = 0.70$ and a response criterion of $c = -0.25$, suggesting measurable sensitivity with a mild bias toward responding ``Yes'', that is, judging a pair as congruent. Balanced accuracy was significantly above chance level, $t(9) = 6.16$, $P < 0.001$, and $d^{\prime}$ was significantly greater than zero, $t(9) = 5.70$, $P < 0.001$.
Accuracy was higher for \textit{congruent pairs} ($M = 72.67\%$, $SD = 12.35\%$) than for \textit{incongruent pairs} ($M = 54.00\%$, $SD = 7.98\%$). A paired-samples $t$-test confirmed this difference, $t(9) = 3.77$, $P = 0.004$, with a large effect size ($d_z = 1.19$) shown in Fig.~\ref{fig:vae_continuity}F.

\section*{Discussion}
This study developed and evaluated a text-conditioned framework for generating GVS waveforms. The evaluation addressed three related questions. 
First, before model training, we examined whether different waveforms in the predefined waveform library could elicit diverse experiences, and whether the distribution of these experiences varied systematically with waveform identity. 
This determines the potential of GVS waveforms as an additional interaction channel, as a narrow range of elicited experiences would limit the information the channel can convey.
Second, we examined whether a limited waveform library could support an organized generative representation. We tested whether latent interpolation produced continuous waveform transitions and whether latent distances covaried with decoded waveform distances.
Third, we tested whether generated waveforms formed discriminable congruent pairings with narrative scenes in unseen participants. Failure to distinguish congruent from incongruent pairings would suggest either insufficient preservation of narrative-relevant waveform features or substantial inter-individual variability in the evoked sensations.

\subsection*{Perceptual diversity and waveform-dependent semantic structure}
The collected descriptions were broadly distributed across the selected semantic-category partition and were not concentrated in a small subset of categories. This distribution suggests that the waveform–description dataset contained a range of reported perceptual and situational associations. Because the category partition was selected to support waveform-level organization analysis, the entropy of this distribution is interpreted here as a descriptive property of the dataset.

However, diversity across the library does not demonstrate waveform-specific perceptual consistency. Because responses were collected through free-form descriptions, variation may partly reflect differences in which aspects of a complex perceptual experience participants chose to emphasize.
The participant-preserving permutation analysis showed that descriptions of the same waveform were more concentrated in similar semantic categories than expected from individual response tendencies alone.
Descriptions associated with the same waveform occupied fewer semantic categories and had a higher proportion of dominant categories. 
Thus, waveform identity influenced the distribution of reported experiences.
This concentration remained limited. On average, each waveform was associated with approximately 8 semantic categories, and the dominant category represented approximately 27\% of the descriptions. Individual waveforms therefore tended to elicit perceptual experiences with some shared structure, while substantial variation remained between participants.

Therefore, from a modeling perspective, GVS waveforms are better treated as stimuli with fuzzy perceptual effects rather than as one-to-one mappings. The evoked sensations are better represented by a multi-label structure than by a single-label assignment, suggesting that retrieval-augmented generation may be better suited than directly mapping discrete labels to waveform features. The goal of waveform generation should therefore be to guide experience toward a target range rather than to produce a specific labeled sensation.

\subsection*{Latent-space continuity and implications for retrieval-based generation}
The analyses indicate that the decoder maps nearby latent representations to relatively similar waveform trajectories and produces gradual changes along sampled interpolation paths.
Small latent perturbations produced only limited waveform changes. Latent interpolation also yielded gradual transitions between decoded waveforms. The interpolation path efficiency was close to the ideal value of \num{1}. In addition, latent distance strongly covaried with decoded waveform distance. Nearby latent representations therefore generally corresponded to similar waveforms.

This organization provides a structural basis for retrieval-based generation and waveform optimization. Local movement in the latent space may support controlled waveform modification and parameter search. It also offers potential for future human-in-the-loop optimization based on user feedback.

\subsection*{Behavioral evidence for narrative--stimulus congruence}
A separate group of participants (n=10) evaluated the waveforms generated by the proposed model. They distinguished congruent from incongruent visual cue--GVS waveform pairings at above-chance levels, as indicated by above-chance balanced accuracy and a significantly positive ($d^{\prime}$). This suggests that the generated waveforms contained perceptual cues relevant to narrative--stimulus matching beyond the participants involved in the initial waveform--text data collection.

However, discriminability remained limited. The overall balanced accuracy was \SI{63.33}{\percent}, with $d^{\prime}=\num{0.70}$. Accuracy was higher for congruent pairings (\SI{72.67}{\percent}) than for incongruent pairings (\SI{54.00}{\percent}), and the negative response criterion ($c=\num{-0.25}$) indicated a mild bias toward congruent judgments. Participants may therefore have found plausible matches easier to identify than clear mismatches. In addition, the observed response bias may have partially reflected demand characteristics. Because participants were not informed of the actual proportion of congruent and incongruent trials, they may have inferred the purpose of the experiment and consequently become more inclined to judge a pair as a "Correct" match rather than "Incorrect" match.

These results support a discriminable matching relationship between generated GVS waveforms and target narratives. However, it does not demonstrate a unique semantic mapping or show that a waveform alone can consistently elicit a specific experience. 

\subsection*{Limitations and future work}
Both the waveform library and the waveform--text dataset remain limited in scale. Because the framework is retrieval-based, its performance depends on the diversity of candidate waveforms and the perceptual information associated with them. Expanding the waveform library and collecting repeated descriptions from more participants would improve the characterization of cross-participant consistency and broaden the retrieval space for natural-language inputs.

For practical applications, the current system may benefit from individualized calibration and closed-loop optimization. Rather than relying solely on group-level waveform--perception relationships, the system could adapt generated waveforms to individual differences in sensitivity.

With continued advances in generative media, particularly in visual and auditory content generation, new forms of multi--modal entertainment built around generated content are likely to emerge. In such applications, visual and auditory content can already be generated automatically, whereas GVS stimuli still largely rely on manually designed waveforms. Methods for automatically generating GVS waveforms from high-level content descriptions therefore remain underdeveloped.
In the absence of such methods, developers must manually design, adjust, and synchronize GVS waveforms for individual scenes or events. This may limit the scalability of GVS-enhanced generative media. The present work addresses this gap by introducing a model that generates GVS waveforms from descriptions of desired sensations. This approach may provide a scalable means of incorporating vestibular cues as an additional sensory modality into future generative entertainment systems, thereby transforming predominantly visual--audio experiences into richer multi--modal experiences.

\section*{Materials and Methods}

\subsection*{Study design}
This study developed and evaluated a text-conditioned framework for GVS waveform generation, with the aim of generating waveforms that are consistent with unstructured descriptions of perceptual sensations under limited data.

The objectives were to determine whether waveform-associated descriptions exhibited non-random semantic organization, 
whether the learned VAE latent space preserved local waveform continuity and global waveform-distance structure,
and whether participants could distinguish congruent from incongruent visual--waveform pairs.

The study consisted of five stages. 
First, we generated and filtered a diverse set of candidate GVS waveforms to construct a waveform library. 
Second, participants experienced the retained waveforms and provided free-text descriptions of their perceived sensations, yielding a waveform--text dataset. 
Third, we encoded the resulting descriptions as dense semantic embeddings using \texttt{intfloat/e5-large-v2}~\cite{Wang2022E5} and clustered them into data-driven semantic categories using K-means~\cite{MacQueen1967}, thereby organizing the waveform--text dataset. 
Fourth, we explored the feasibility of natural-language-conditioned candidate waveform generation using a retrieval-augmented VAE framework based on prior methods. The framework generated candidate waveforms from target text descriptions through semantic retrieval, latent-posterior fusion, and latent-space optimization. 
Finally, we constructed congruent and incongruent visual--waveform pairs and evaluated them using a two-choice congruence-judgment task~\cite{Ratcliff2018}.

\subsection*{Participants and GVS setup}

\begin{figure}[ht]
    \centering
    \includegraphics[width=0.9\linewidth]{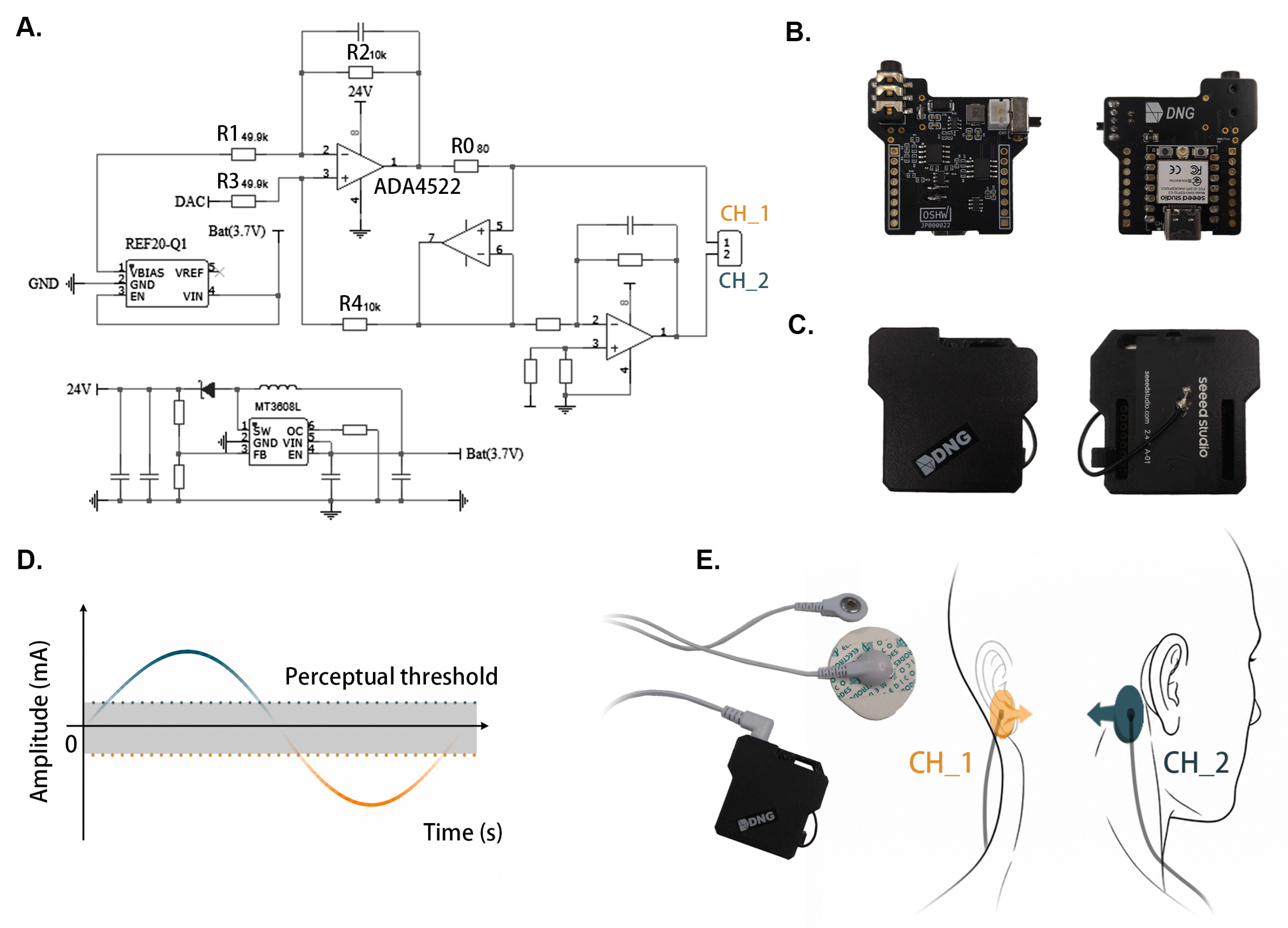}
    \caption{Hardware architecture and electrode configuration of the wearable dual-channel 
    GVS module.
    (A) Circuit schematic of the digital-to-analog converter (DAC)-controlled dual-channel stimulation-output stage and the associated power-regulation circuit. Channels $\mathrm{CH}_1$ and $\mathrm{CH}_2$ provide independent stimulation outputs.
    (B) Front and rear views of the custom printed circuit board (PCB).
    (C) Front and rear views of the assembled wearable enclosure.
    (D) Illustration of a biphasic stimulation waveform. Positive and negative waveform excursions are shown in teal and orange, respectively, relative to the perceptual-threshold reference indicated by the dotted lines.
    (E) Electrode connections and bilateral placement over the mastoid regions. The orange and teal electrodes correspond to $\mathrm{CH}_1$ and $\mathrm{CH}_2$, respectively.}
    \label{fig:prototype}
\end{figure}

\subsubsection*{Participants and ethics}
Sixteen participants completed the waveform--description pretest, during which they evaluated the retained GVS waveform library and provided verbal descriptions of their perceived sensations. This pretest group included 3 female and 13 male participants, with ages ranging from 22 to 32 years (M = 26.2, SD = 2.59).
The resulting waveform--description pairs formed the waveform--text dataset used to train the text-conditioned waveform-generation framework and evaluate waveform-level semantic organization.

An independent group of 10 participants completed the subsequent behavioral evaluation, in which they performed the two-choice congruence-judgment task. This validation group included 4 female and 6 male participants, with ages ranging from 22 to 30 years (M = 26.8, SD = 2.66).

\subsubsection*{GVS hardware and calibration}
The dual-channel wearable GVS module used in this study was modified from an open-source GVS device \cite{Liu2025}. It implemented a voltage-controlled enhanced Howland current source (EHCS). Based on the safety range of transcranial electrical stimulation reported in \cite{Safety}, the target output range was defined as \SI{-4}{\milli\ampere} to \SI{4}{\milli\ampere}. In this module, the output current was determined by Equation~\ref{eq:output-current}:
\begin{equation}
\frac{R_1}{R_2} = \frac{R_3}{R_4} = k,
\qquad
i_L = \frac{u_{\mathrm{in}+} - u_{\mathrm{in}-}}{kR_0}.
\label{eq:output-current}
\end{equation}
As shown in Fig.~\ref{fig:prototype}A,  \(u_{\mathrm{in}-}\) was fixed at \SI{1.65}{\volt}, whereas \(u_{\mathrm{in}+}\) ranged from \SI{0}{\volt} to \SI{3.3}{\volt}. An additional \SI{0.125}{\milli\ampere} margin was reserved beyond each end of this range for calibration and error compensation. The scaling resistance \(R_0\) was adjusted to \SI{80}{\ohm}, resulting in an output range of approximately \SI{-4.125}{\milli\ampere} to \SI{4.125}{\milli\ampere}.
Before experimentation, each prototype was measured and calibrated using piecewise-linear interpolation to compensate for output-current deviations. By restricting the calibrated digital-to-analog converter range, in which digital codes from 0 to 4095 corresponded to \SI{0}{\volt} to \SI{3.3}{\volt}, the maximum operating current of each module was constrained to \SI{-4}{\milli\ampere} to \SI{4}{\milli\ampere} between \(\mathrm{CH}_1\) and \(\mathrm{CH}_2\), without introducing a current bias.

A Seeeduino ESP32C3 microcontroller unit (MCU) supported power management and wireless communication with the host device. The combined printed circuit board measured approximately \(42 \times 39~\mathrm{mm}\) (Fig.~\ref{fig:prototype}B). After integration with the three-dimensional printed protective enclosure, the overall device dimensions were approximately \(43 \times 43 \times 20~\mathrm{mm}\) (Fig.~\ref{fig:prototype}C).

The GVS module was remotely controlled by a PC-based Unity application via Wi-Fi. Waveform data were transmitted as user datagram protocol (UDP) messages to the MCU, which controlled the MCP4725 digital-to-analog converter (DAC) module to drive current control (Fig.~\ref{fig:prototype}D).

Before stimulation, electrode sites were prepared using Nuprep Skin Prep Gel (Weaver and Company, Houston, TX, USA) to reduce electrode--skin impedance. Gel was then fully removed, and MSGST-37 ECG electrodes were affixed over the mastoid regions (Fig.~\ref{fig:prototype}E).

\subsubsection*{Output verification}
Electrical output was evaluated by measuring the differential voltage across load resistors representing a range of inter-mastoid impedances. The measured resistances were \SI{984}{\ohm}, \SI{2167}{\ohm}, and \SI{3840}{\ohm}, as measured using a Sanwa CD771 multimeter. 
Each resistance condition included the 100 retained experimental waveforms and two additional triangular calibration waveforms spanning the maximum output range (\SI{-4}{\milli\ampere} to \SI{4}{\milli\ampere}).

Current measurements were obtained using a PicoScope 5244D operated with PicoScope 7 T\&M software. The oscilloscope was configured for differential voltage measurement using Channel A and B, each set to a \SI{20}{\volt} range and 15-bit resolution. The sampling rate was \(10\,\mathrm{kS/s}\).

\subsection*{Waveform--Text Dataset and Semantic Processing}

\subsubsection*{Candidate waveform construction and filtering}
A total of 500 candidate stimulation waveforms were generated for the perceptual-labeling task. Each waveform file contained 250 samples corresponding to a waveform \SI{5}{\second} sampled at \SI{50}{\hertz}.

\begin{figure}[ht]
    \centering
    \includegraphics[width=1\linewidth]{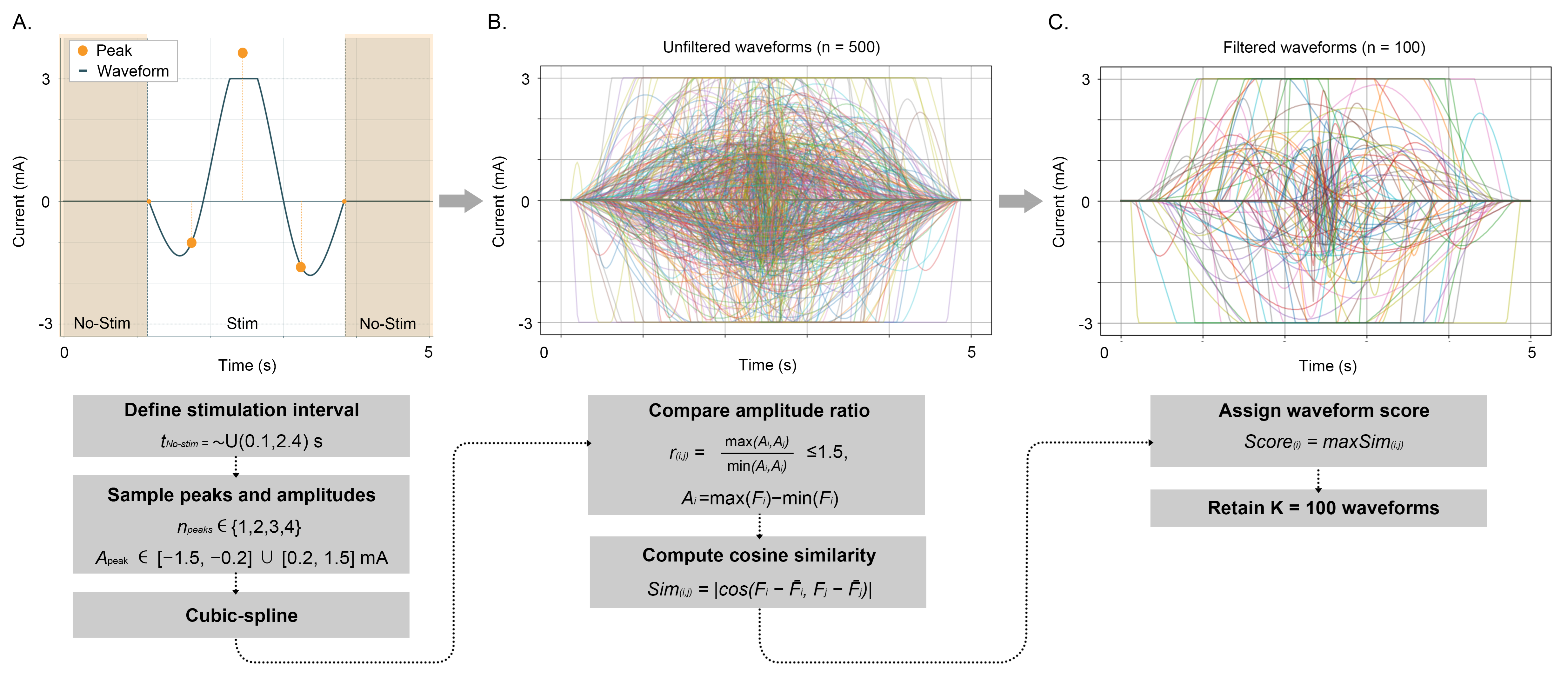}
    \caption{Generation and diversity-based filtering of candidate stimulation waveforms.
(A) Candidate waveforms were generated by sampling the duration of the no-stimulation intervals, $t_{\mathrm{No\text{-}stim}} \sim \mathcal{U}(0.1, 2.4)\,\mathrm{s}$, the number of peaks, $n_{\mathrm{peaks}} \in \{1,2,3,4\}$, and the peak amplitudes, $A_{\mathrm{peak}} \in [-1.5,-0.2] \cup [0.2,1.5]\,\mathrm{mA}$. Cubic-spline interpolation was then applied to obtain continuous waveforms. Orange markers indicate sampled peaks, and shaded regions indicate no-stimulation intervals. 
(B) A total of $n=500$ unfiltered waveforms were screened using the pairwise peak-to-peak amplitude ratio, $r_{ij} = \max(A_i,A_j)/\min(A_i,A_j) \leq 1.5$, where $A_i = \max(F_i)-\min(F_i)$ and $F_i$ denotes waveform $i$. Cosine similarity was subsequently calculated between mean-centered waveform vectors. 
(C) Each waveform was assigned a similarity score, $\mathrm{Score}_i = \max_j \mathrm{Sim}_{ij}$, and the $K=100$ waveforms with the lowest scores were retained to reduce waveform redundancy. Colored traces represent individual waveforms.}
    \label{fig:Gen}
\end{figure}

A zero-amplitude no-stimulation interval was placed at both the beginning and end of each waveform, with the same duration at both ends. This duration was sampled between \SI{0.1}{\second} and \SI{2.4}{\second}, thereby defining the central stimulation interval.
Within the stimulation interval, one to four peak points were sampled, with probabilities of \SI{35}{\percent}, \SI{35}{\percent}, \SI{25}{\percent}, and \SI{5}{\percent}, respectively. For each peak, positive and negative polarities were selected with equal probability, and the peak amplitude was sampled between \SI{0.2}{\milli\ampere} and \SI{1.5}{\milli\ampere} magnitude. 
The start and end points of the stimulation interval were fixed at zero amplitude. Cubic-spline interpolation was then applied to the sampled peaks and boundary points to generate continuous waveforms. To account for user comfort, waveform amplitudes were clipped to \SI{-3}{\milli\ampere} to \SI{3}{\milli\ampere} (Fig.~\ref{fig:Gen}A).

The 500 candidate waveforms were filtered to reduce redundancy and retain a diverse waveform library (Fig.~\ref{fig:Gen}B). For each waveform, peak-to-peak amplitude was calculated as the difference between its maximum and minimum current values. Pairwise comparisons were restricted to waveforms whose peak-to-peak amplitude ratio did not exceed 1.5.

The cosine similarity was then calculated between the mean-centered waveform vectors. To prevent retention of waveforms that differed only in polarity, each comparison also considered the polarity-inverted version of the corresponding waveform. For each candidate waveform, the highest similarity observed across all eligible comparison waveforms was used as its redundancy score. The 100 waveforms with the lowest redundancy scores were retained (Fig.~\ref{fig:Gen}C).
\label{sec:gui_validation}

\subsubsection*{GUI-based waveform delivery and validation procedure}
\begin{figure}[h]
    \centering
    \includegraphics[width=1\linewidth]{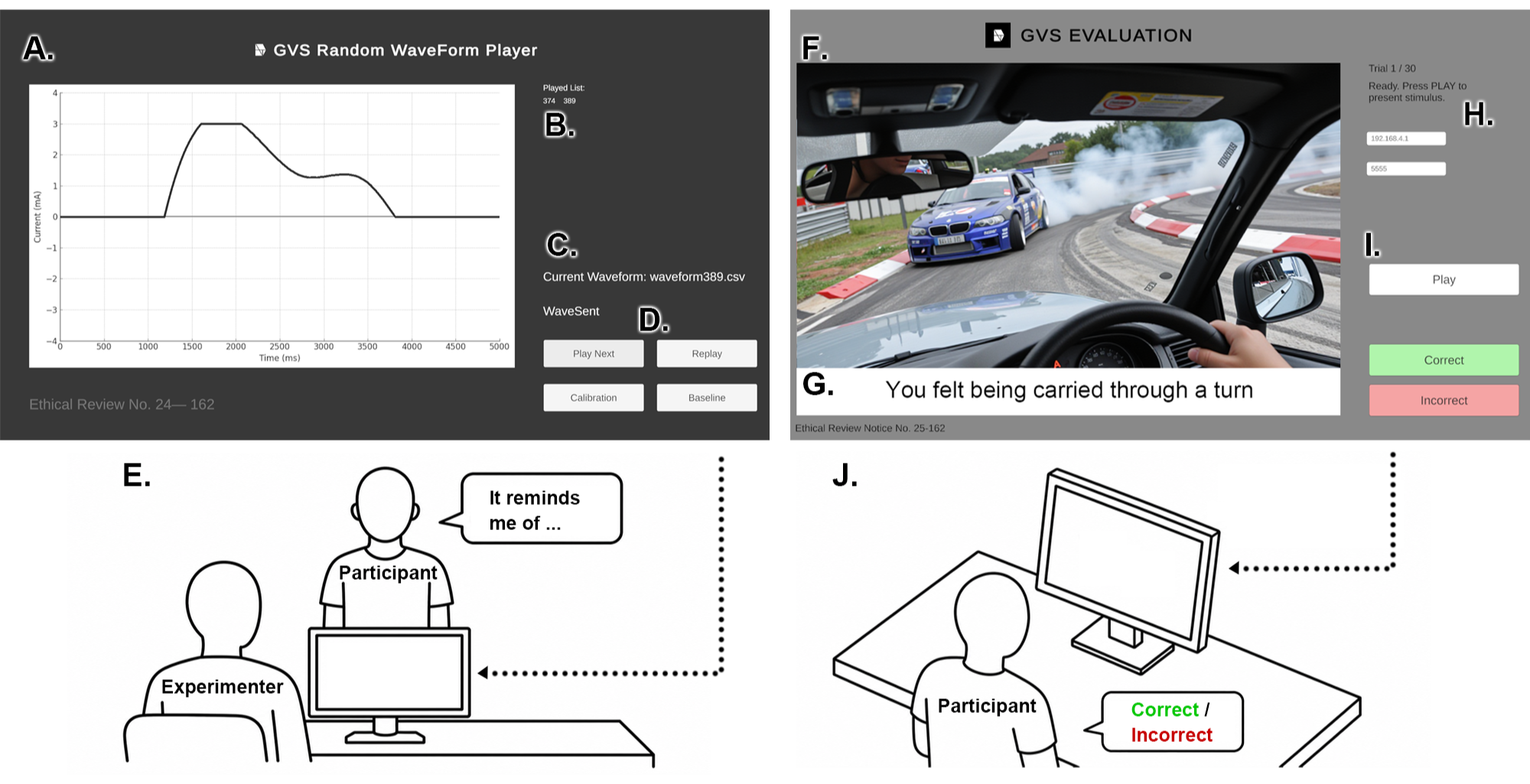}
    \caption{GUI and experimental procedure for GVS waveform delivery and evaluation.
(A) The GVS Random WaveForm Player displays the stimulation waveform as current (mA) over time (ms). 
(B) The playlist lists identifiers of previously delivered waveforms. 
(C) The status panel displays the active waveform file and its transmission status. 
(D) Playback controls allow the experimenter to present the next randomly selected waveform, replay the preceding waveform, or deliver calibration and baseline waveforms. 
(E) Following stimulation, the experimenter obtains a verbal description of the participant's perceived sensation. 
(F) The GVS Evaluation interface presents a visual stimulus during the evaluation trial. 
(G) A textual description associated with the visual stimulus is displayed beneath the image. 
(H) Network settings specify the connection to the GVS system. 
(I) Controls initiate stimulus delivery and allow participants to complete a two-choice congruence-judgment by selecting ``Correct'' or ``Incorrect.''
(J) The participant views the evaluation stimulus on a monitor and provides a correct--incorrect judgment.}
\label{fig:gvs_interfaces_and_evaluation}
\end{figure}

Waveform--text acquisition and subsequent validation were controlled using a Unity-based graphical user interface (GUI; Fig.~\ref{fig:gvs_interfaces_and_evaluation}A--J). The GUI consisted of two functional parts. Panels A--D supported the waveform--text acquisition phase, during which the experimenter delivered GVS waveforms and collected participants' verbal descriptions of the perceived sensations. 
Panels F--G supported the validation phase, during which visual stimuli derived from the waveform--text records were presented, and participants judged whether the evoked sensations were congruent with the presented visual stimuli.

In the waveform--text acquisition phase, waveform CSV files were loaded from a local waveform library at the beginning of each session. The files were then randomly shuffled into a playback queue without replacement. During each presentation, the selected waveform was visualized in the Unity interface for experimenter monitoring (Fig.~\ref{fig:gvs_interfaces_and_evaluation}A). The GUI was shown only to the experimenter, not to the participant.
The playlist showed identifiers of waveforms that had already been delivered (Fig.~\ref{fig:gvs_interfaces_and_evaluation}B). The status panel indicated the active waveform file and its transmission status (Fig.~\ref{fig:gvs_interfaces_and_evaluation}C). Calibration and baseline waveforms were handled separately from the randomized queue. These waveforms were used to verify waveform delivery before each experiment (Fig.~\ref{fig:gvs_interfaces_and_evaluation}D).

The retained 100 waveforms were presented to 16 participants. Trials were self-paced. After each waveform, the participant provided a verbal description of the perceived sensation (Fig.~\ref{fig:gvs_interfaces_and_evaluation}E). The next waveform was presented only after this response had been recorded. 
A scheduled rest break was provided after 50 waveform presentations. The duration of the break was determined by the participant. Participants could stop the experiment at any time, request additional breaks, or ask for a waveform to be replayed before finalizing their description. No waveform was repeated as part of the randomized playback queue except for such participant-requested replays.
Of the 1,600 waveform--text records, placeholder responses and missing-data entries were excluded before analysis. The resulting dataset contained 1,526 valid waveform--text records.

The validation phase was conducted with a separate group of 10 new participants who had not taken part in the waveform--text acquisition phase. 
In this phase, the evaluation interface presented the visual cue to the participant before waveform delivery (Fig.~\ref{fig:gvs_interfaces_and_evaluation}F). The associated textual description was displayed beneath the visual cue (Fig.~\ref{fig:gvs_interfaces_and_evaluation}G). These elements remained on the screen throughout the trial. This allowed the participant to compare the displayed cue with the sensation evoked by the delivered waveform.
The side panel showed trial information and network settings for communication with the GVS delivery system (Fig.~\ref{fig:gvs_interfaces_and_evaluation}H). The response controls were used to initiate waveform delivery, replay the waveform on request, and record the participant's two-choice congruence judgment as ``Correct'' or ``Incorrect'' (Fig.~\ref{fig:gvs_interfaces_and_evaluation}I). During the trial, the participant viewed the evaluation stimulus on a monitor and submitted the judgment after experiencing each waveform (Fig.~\ref{fig:gvs_interfaces_and_evaluation}J).
\label{sec:gui_validation2}

\subsubsection*{Semantic embedding, clustering, and prompts}
Sentence embeddings were calculated for all valid descriptions using \texttt{intfloat/e5-large-v2}. To reduce participant-specific response-style effects during semantic clustering, each participant's mean embedding, calculated across that participant's valid responses, was subtracted from each of their response embeddings. The resulting centered embeddings were vector-normalized before clustering.

Descriptions reflecting low-intensity or unclear sensations, including descriptions such as ``weak,'' ``unnoticeable,'' and ``hard to describe,'' were assigned to a predefined low-intensity category. The remaining descriptions were assigned automatically to 14 data-driven clusters using \(k\)-means clustering. Together, the 14 data-driven clusters and the predefined low-intensity category yielded 15 semantic categories. Representative cluster labels were assigned by inspecting representative descriptions within each category and were used for descriptive presentation only.

The number of data-driven clusters was fixed at \(k=14\) such that the resulting 15-category partition was comparable to the mean number of valid descriptions available per waveform (\(1{,}526/100\)). This choice provided an interpretable range for waveform-level semantic-organization analyses: using substantially fewer categories would constrain the maximum number of categories that could be represented for a given waveform, whereas a substantially larger category set would yield sparse waveform-by-category counts.

Under the permutation null hypothesis, semantic-category assignments are independent of waveform identity while the empirical category structure is preserved. Thus, descriptions associated with an individual waveform may be distributed across multiple semantic categories, with the expected number of represented categories and dominant-category proportion determined empirically from the permutation distribution. In contrast, waveform-specific semantic organization would be indicated by fewer represented categories per waveform and greater concentration of descriptions within a dominant category. The final 15-category partition was therefore selected to provide sufficient dynamic range for the planned permutation-based analyses of waveform-level semantic organization.

\subsection*{Text-Driven Waveform Generation}
\subsubsection*{Retrieval-guided waveform generation}
We implemented a retrieval-guided procedure to generate stimulation waveforms
from natural-language descriptions. The procedure combines
semantic retrieval with sensitivity-aware fusion of VAE latent posteriors.

A retrieval database was constructed by encoding each stimulation
waveform with the Conv1D-VAE and each associated textual description with
E5-large-v2. The database stored text embeddings and waveform-associated latent posterior parameters.
For a new textual input, its embedding was generated locally using E5-large-v2 and compared with precomputed text embeddings stored in the local database to
retrieve the top-$K$ most similar records. For each retrieved record,
the cosine similarity score \(s_i\) and the associated latent posterior
parameters \((\boldsymbol{\mu}_i,\log\boldsymbol{\sigma}_i^2)\) were obtained.
The posterior mean \(\boldsymbol{\mu}_i\) was decoded to generate a candidate
waveform. Candidate waveforms were adjusted to the requested direction when
applicable and sign-aligned to the top-ranked candidate before waveform-domain
fusion.

The local latent sensitivity (LLS) was calculated around each retrieved latent
mean. Quantified the change in the decoded waveform after small
perturbations in latent space (Equation~\ref{eq:LLS}):

\begin{equation}
\mathrm{LLS}_i
=
\frac{1}{M}
\sum_{m=1}^{M}
\frac{
\left\|
g_\theta(\boldsymbol{\mu}_i+\boldsymbol{\delta}_{im})
-
g_\theta(\boldsymbol{\mu}_i)
\right\|_2
}{
\left\|
\boldsymbol{\delta}_{im}
\right\|_2
},
\label{eq:LLS}
\end{equation}
where \(\boldsymbol{\delta}_{im}\) denotes the \(m\)-th small random
perturbation applied to the latent mean of the \(i\)-th retrieved candidate.
Lower LLS values indicated smaller waveform changes around the corresponding
latent mean.

Fusion scores combined semantic similarity, LLS, and
mean posterior log-variance (Equation~\ref{eq:fusion}):

\begin{equation}
a_i
=
s_i
-
\beta\,\mathrm{LLS}_i
+
\gamma\,\overline{\log\boldsymbol{\sigma}_i^2},
\qquad
w_i
=
\frac{\exp(a_i/\tau)}
{\sum_{j\in\mathcal{I}_K}\exp(a_j/\tau)}.
\label{eq:fusion}
\end{equation}

Here, \(a_i\) is the unnormalized fusion score of the \(i\)-th retrieved
candidate, and \(w_i\) is its normalized fusion weight. The term \(s_i\)
denotes cosine similarity between the input text and the \(i\)-th retrieved
description. The term
\(\overline{\log\boldsymbol{\sigma}_i^2}\) denotes the mean posterior
log-variance across latent dimensions. The coefficients \(\beta\) and
\(\gamma\) control the contributions of LLS and posterior log-variance,
respectively, whereas \(\tau\) controls the concentration of the softmax
weights.

The fusion weights were used to construct a waveform-domain target from the
aligned candidate waveforms and to combine the retrieved latent posteriors.
The fused posterior \(F\) was approximated by moment matching (Equation~\ref{eq:MomentM}):

\begin{equation}
\begin{aligned}
\boldsymbol{\mu}_{\mathrm{F}}
&=
\sum_{i\in\mathcal{I}_K}
w_i\boldsymbol{\mu}_i, \\
\boldsymbol{\sigma}_{\mathrm{F}}^2
&=
\sum_{i\in\mathcal{I}_K}
w_i
\left[
\boldsymbol{\sigma}_i^2
+
\left(
\boldsymbol{\mu}_i-
\boldsymbol{\mu}_{\mathrm{F}}
\right)^2
\right].
\end{aligned}
\label{eq:MomentM}
\end{equation}

The fused posterior \(F\) defines a local latent region centered at
\(\boldsymbol{\mu}_{\mathrm{F}}\), with a spread described by
\(\boldsymbol{\sigma}_{\mathrm{F}}^2\). The target in the waveform-domain specifies 
the desired waveform morphology, while \(F\) specifies the latent region
used for subsequent optimization.

The fused posterior and waveform-domain targets were then used to guide
latent-space optimization. The resulting decoded waveform was subsequently
passed to the post-processing stage.

\subsubsection*{Latent optimization and post-processing}
Latent optimization was used to obtain a decoder-compatible waveform that
closely matched the fused waveform-domain target. Post-processing was applied
to preserve the polarity and amplitude scale of semantically matched
candidates and to suppress high-frequency artifacts in the VAE-decoded
waveform.

Latent initializations were sampled around the fused posterior and optimized
against the fused waveform-domain target. The solution with the lowest
recorded optimization loss was selected and decoded. The decoded waveform was
sign-aligned to the top-ranked retrieved candidate and peak-matched to the
mean absolute peak magnitude of the retrieved candidates. A
Savitzky--Golay filter was then applied, followed by sign alignment and peak
matching, to produce the final output. 

\begin{figure}[ht]
    \centering
    \includegraphics[width=1\linewidth]{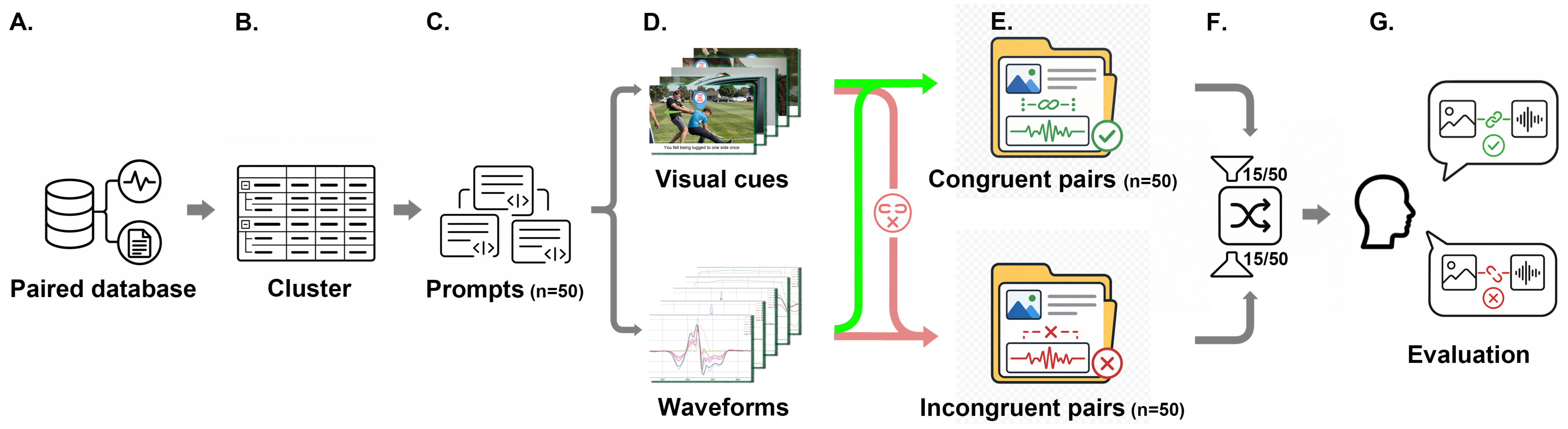}
\caption{Experimental workflow for constructing and evaluating visual--waveform stimulus pairs. 
(A) The waveform--text dataset comprising 1,526 waveform--description records was assembled. 
(B) The dataset was organized into semantic categories. 
(C) Fifty textual prompts were generated from the resulting semantic categories. 
(D) Each prompt was used to generate corresponding visual cues and waveforms. 
(E) Visual cues and waveforms generated from the same semantic category were combined to form 50 congruent pairs. Incongruent pairs were constructed by randomly pairing visual cues and waveforms generated from different semantic categories while excluding within-category pairings, yielding 50 incongruent pairs. Green check marks and red crosses indicate congruent and incongruent pairings, respectively. 
(F) For each participant, 15 pairs from each condition were randomly sampled and presented in a randomized order. 
(G) Participants completed a binary congruence-judgment task, classifying each visual--waveform pair as congruent or incongruent.}
    \label{fig:stimulus_pairing_workflow}
\end{figure}

\subsection*{Stimulus Pair Construction for Behavioral Validation}
The waveform--text dataset, was first assembled (Fig.~\ref{fig:stimulus_pairing_workflow}A) and clustered into 10 semantic categories for stimulus construction, with consideration given to reduce the total trial number, and the need to assign an equal number of prompts to each category (Fig.~\ref{fig:stimulus_pairing_workflow}B). This partition was separate from the 15-category solution used for waveform-level statistical analysis. 
Fifty textual prompts were then generated from the 10 semantic categories (Fig.~\ref{fig:stimulus_pairing_workflow}C). Visual--waveform stimulus pairs were constructed for behavioral validation, with an equal number of pairs assigned to each category.
For each prompt, a combined image--text visual cue and a corresponding waveform were
generated (Fig.~\ref{fig:stimulus_pairing_workflow}D). The visual components
were generated using Z-Image-Turbo\cite{zimage2025} at a resolution of \(2164 \times 1400\)
pixels, whereas each waveform was generated from its corresponding textual
prompt using the retrieval-guided waveform-generation framework described
above.

Visual cues and waveforms originating from the same semantic category were combined to form 50 congruent pairs. Incongruent pairs were constructed by randomly combining visual cues and waveforms from different semantic categories, while excluding within-category pairings, yielding 50 incongruent pairs
(Fig.~\ref{fig:stimulus_pairing_workflow}E). Individual visual cues and
waveforms could appear in more than one pair across the stimulus set.
Before the main validation task, participants completed three practice sets.
For each validation participant, 15 congruent and 15 incongruent pairs were
randomly sampled and presented in randomized order, yielding 30 experimental
trials (Fig.~\ref{fig:stimulus_pairing_workflow}F). Trial progression was
self-paced, with no fixed fixation period, response deadline, inter-trial
interval, or scheduled rest breaks. Behavioral responses were recorded as
two-choice congruence judgments for subsequent analysis
(Fig.~\ref{fig:stimulus_pairing_workflow}G).
\label{sec:stimulus_pair_construction}

\subsection*{Ethical Approval}
This study was conducted in accordance with the ethical guidelines established by the Department of Informatics, University of Tsukuba. The nature and procedures of the study were explained to all participants before participation, and written informed consent was obtained before the study began. Participants were informed of their right to withdraw at any time without consequences.

\subsection*{Statistical Analysis}
Statistical analyses were conducted to evaluate waveform-specific semantic organization, latent--waveform geometric consistency, and behavioral discriminability. Statistical significance was assessed at an alpha level of \(0.05\).

Waveform-specific semantic organization was evaluated using a participant-preserving permutation test. Waveform assignments were independently shuffled within each participant, thereby preserving participant-specific response and semantic-category distributions. A total of 10,000 permutations were performed. The observed mean number of unique semantic categories represented per waveform and mean dominant-category proportion were compared with their respective permutation null distributions. Two-sided empirical \(P\) values were calculated using a plus-one correction.

Latent--waveform geometric consistency was quantified using Pearson's correlation coefficient, \(r\). We randomly sampled 1,000 pairs of distinct latent representations and compared the Euclidean distance between their VAE posterior means with the RMS distance between the corresponding decoded waveforms. Pearson's \(r\) was used as a descriptive measure of correspondence between latent-space geometry and decoded waveform variation.

Behavioral analyses were conducted at the participant level for the independent validation group (\(N = 10\)). Congruent and incongruent pairs were treated as signal-present and signal-absent trials, respectively, and hit rate, false-alarm rate, sensitivity (\(d'\)), response criterion (\(c\)), and balanced accuracy were calculated. Log-linear correction was applied for signal-detection indices. Balanced accuracy and \(d'\) were tested using one-sided one-sample \(t\)-tests against 0.5 and 0, respectively, whereas congruent- and incongruent-pair accuracies were compared using a two-sided paired-samples \(t\)-test with Cohen's \(d_z\) as the effect size. Statistical analyses were performed in Python using NumPy and SciPy.

\section*{Acknowledgments}
The authors thank all participants in the dataset-collection and behavioral-validation studies for their time, cooperation, and valuable feedback.

\subsection*{Author Contributions}
Z. Liu, T. Fushimi, and Y. Ochiai conceived the idea and designed the study. Z. Liu conducted the experiments, collected the data, performed the data analysis, developed the hardware and software system, implemented the waveform-generation framework, and prepared the visualizations. Z. Liu wrote the original draft of the manuscript. T. Fushimi and Y. Ochiai supervised the project and contributed to the interpretation of the results. All authors reviewed and approved the final manuscript.

\subsection*{Funding}
This work was supported by Pixie Dust Technologies, Inc.

\subsection*{Competing Interests}
The authors declare that they have no competing interests.

\subsection*{AI Tool Usage Declaration}
During the conduct of this study and the preparation of the manuscript, the authors used ChatGPT for language refinement and Z-Image-Turbo to generate visual stimuli used in the experiments. All AI-assisted content and AI-generated images were subsequently reviewed and verified by the authors, who take full responsibility for the final content of the manuscript.

\section*{DATA AVAILABILITY}
The final generative model and its source code, the data used in the validation experiments---including the waveform-generation parameters associated with each generated waveform, the corresponding visual guidance, and the GVS waveforms used for evaluation---and the model-validation data derived from user judgments are provided as supplementary materials on Zenodo (10.5281/zenodo.21670229).

Any data that could potentially identify individual participants are provided only in de-identified form, in accordance with the conditions of informed consent and applicable institutional ethical requirements. No materials used in this study are subject to a Material Transfer Agreement.

\printbibliography[title={REFERENCES}]

@ARTICLE{Liu2025,
title = {Design and evaluation of a voltage-controlled current source for galvanic vestibular stimulation research},
journal = {HardwareX},
volume = {22},
pages = {e00647},
year = {2025},
issn = {2468-0672},
doi = {https://doi.org/10.1016/j.ohx.2025.e00647},
author = {Zhi Liu and Shieru Suzuki and Tatsuki Fushimi and Yoichi Ochiai}
}

@article{FrontiersGVSreview,
  author = {Valter, Yishai and Vataksi, Linda and Allred, Aaron R.
            and Hebert, Jeffrey R. and Brunyé, Tad T.
            and Clark, Torin K. and Serrador, Jorge
            and Datta, Abhishek},
  title = {A review of parameter settings for {Galvanic Vestibular
           Stimulation} in clinical applications},
  journal = {Frontiers in Human Neuroscience},
  year = {2025},
  volume = {19},
  pages = {1518727},
  doi = {10.3389/fnhum.2025.1518727},
  issn = {1662-5161}
}

@inproceedings{balanceNinja,
  author = {Byrne, Richard and Marshall, Joe and Mueller, Florian},
  title = {Balance Ninja: Towards the Design of Digital Vertigo Games
           via Galvanic Vestibular Stimulation},
  booktitle = {Proceedings of the 2016 Annual Symposium on
               Computer-Human Interaction in Play},
  year = {2016},
  pages = {159--170},
  numpages = {12},
  publisher = {Association for Computing Machinery},
  address = {New York, NY, USA},
  isbn = {9781450344562},
  doi = {10.1145/2967934.2968080},
  location = {Austin, Texas, USA},
  series = {CHI PLAY '16}
}

@inproceedings{rollerCoaster,
author = {Aoyama, Kazuma and Higuchi, Daiki and Sakurai, Kenta and Maeda, Taro and Ando, Hideyuki},
title = {GVS RIDE: providing a novel experience using a head mounted display and four-pole galvanic vestibular stimulation},
year = {2017},
isbn = {9781450350129},
publisher = {Association for Computing Machinery},
address = {New York, NY, USA},
doi = {10.1145/3084822.3084840},
booktitle = {ACM SIGGRAPH 2017 Emerging Technologies},
articleno = {9},
numpages = {2},
location = {Los Angeles, California},
series = {SIGGRAPH '17}
}

@ARTICLE{GVSbike,
author={Groth, Colin and Tauscher, Jan-Philipp and Heesen, Nikkel and Hattenbach, Max and Castillo, Susana and Magnor, Marcus},
journal={ IEEE Transactions on Visualization \& Computer Graphics },
title={{ Omnidirectional Galvanic Vestibular Stimulation in Virtual Reality }},
year={2022},
volume={28},
number={05},
ISSN={1941-0506},
pages={2234-2244},
doi={10.1109/TVCG.2022.3150506},
publisher={IEEE Computer Society},
address={Los Alamitos, CA, USA},
month=may}

@inproceedings{RedirectionW,
author = {Hwang, Seokhyun and Lee, Jieun and Kim, Youngin and Seo, Youngseok and Kim, Seungjun},
title = {Electrical, Vibrational, and Cooling Stimuli-Based Redirected Walking: Comparison of Various Vestibular Stimulation-Based Redirected Walking Systems},
year = {2023},
isbn = {9781450394215},
publisher = {Association for Computing Machinery},
address = {New York, NY, USA},
doi = {10.1145/3544548.3580862},
booktitle = {Proceedings of the 2023 CHI Conference on Human Factors in Computing Systems},
articleno = {767},
numpages = {18},
location = {Hamburg, Germany},
series = {CHI '23}
}

@inproceedings{RedirectHand,
author = {Katori, Kensuke and Tanaka, Yudai and Ochiai, Yoichi and Lopes, Pedro},
title = {Vestibular Stimulation Enhances Hand Redirection},
year = {2025},
isbn = {9798400720376},
publisher = {Association for Computing Machinery},
address = {New York, NY, USA},
doi = {10.1145/3746059.3747776},
booktitle = {Proceedings of the 38th Annual ACM Symposium on User Interface Software and Technology},
articleno = {62},
numpages = {10},
location = {
},
series = {UIST '25}
}

@article{fewshot,
  title={A Survey on Generative Modeling with Limited Data, Few Shots, and Zero Shot},
  author={Abdollahzadeh, Milad and Liu, Guimeng and Malekzadeh, Touba and Teo, Christopher T.H and Chandrasegaran, Keshigeyan and Cheung, Ngai-Man},
  journal={Transactions on Machine Learning Research},
  year={2025},
  url={https://openreview.net/forum?id=u7GTHazuRp},
  note={Survey Certification}
}

@misc{ttscgan,
  title={TTS-CGAN: A Transformer Time-Series Conditional GAN for Biosignal Data Augmentation},
  author={Li, Xiaomin and Ngu, Anne Hee Hiong and Metsis, Vangelis},
  year={2022},
  eprint={2206.13676},
  archivePrefix={arXiv},
  primaryClass={cs.LG},
  doi={10.48550/arXiv.2206.13676},
}

@InProceedings{VerbalTS,
  title = {{V}erbal{TS}: Generating Time Series from Texts},
  author = {Gu, Shuqi and Li, Chuyue and Jing, Baoyu and Ren, Kan},
  booktitle = {Proceedings of the 42nd International Conference on Machine Learning},
  pages = {20448--20476},
  year = {2025},
  editor = {Singh, Aarti and Fazel, Maryam and Hsu, Daniel and Lacoste-Julien, Simon and Berkenkamp, Felix and Maharaj, Tegan and Wagstaff, Kiri and Zhu, Jerry},
  volume = {267},
  series = {Proceedings of Machine Learning Research},
  month = {13--19 Jul},
  publisher = {PMLR},
  url = {https://proceedings.mlr.press/v267/gu25a.html}
}

@article{LRtilt,
  author  = {Wardman, D. L. and Day, B. L. and Fitzpatrick, R. C.},
  title   = {Position and velocity responses to galvanic vestibular stimulation in human subjects during standing},
  journal = {The Journal of Physiology},
  year    = {2003},
  volume  = {547},
  number  = {1},
  pages   = {293--299},
  doi     = {10.1113/jphysiol.2002.030767}
}

@article{Demura2006,
  author  = {Demura, Shinichi and Kitabayashi, Tamotsu},
  title   = {Comparison of Power Spectrum Characteristics of Body Sway during a Static Upright Standing Posture in Healthy Elderly People and Young Adults},
  journal = {Perceptual and Motor Skills},
  year    = {2006},
  volume  = {102},
  number  = {2},
  pages   = {467--476},
  doi     = {10.2466/pms.102.2.467-476}
}

@article{Dizzness,
author = {Carmona, Sergio and Ferrero, Antonela and Pianetti, Guillermina and Escolá, Natalia and Arteaga, María Victoria and Frankel, Lilian},
title = {Galvanic vestibular stimulation improves the results of vestibular rehabilitation},
journal = {Annals of the New York Academy of Sciences},
volume = {1233},
number = {1},
pages = {E1-E7},
doi = {https://doi.org/10.1111/j.1749-6632.2011.06269.x},
year = {2011}
}

@article{Parkinson,
  author = {Lee, Soojin and Smith, Paul F. and Lee, Won Hee
            and McKeown, Martin J.},
  title = {Frequency-Specific Effects of Galvanic Vestibular Stimulation
           on Response-Time Performance in {Parkinson's Disease}},
  journal = {Frontiers in Neurology},
  year = {2021},
  volume = {12},
  pages = {758122},
  doi = {10.3389/fneur.2021.758122},
  issn = {1664-2295}
}

@inproceedings{RAGDefine,
  title     = {Retrieval-Augmented Generation for Knowledge-Intensive NLP Tasks},
  author    = {Lewis, Patrick and Perez, Ethan and Piktus, Aleksandra and Petroni, Fabio and Karpukhin, Vladimir and Goyal, Naman and K{\"u}ttler, Heinrich and Lewis, Mike and Yih, Wen-tau and Rockt{\"a}schel, Tim and Riedel, Sebastian and Kiela, Douwe},
  booktitle = {Advances in Neural Information Processing Systems},
  volume    = {33},
  pages     = {9459--9474},
  year      = {2020}
}

@article{GVSutz2011minor,
  title={Minor adverse effects of galvanic vestibular stimulation in persons with stroke and healthy individuals},
  author={Utz, Kathrin S. and Korluss, Kathia and Schmidt, Lena and Rosenthal, Alisha and Oppenl{\"a}nder, Karin and Keller, Ingo and Kerkhoff, Georg},
  journal={Brain Injury},
  volume={25},
  number={11},
  pages={1058--1069},
  year={2011},
  doi={10.3109/02699052.2011.607789}
}

@article{GVSgoel2015,
  title={Using low levels of stochastic vestibular stimulation to improve balance function},
  author={Goel, Rajan and Kofman, Igor and Jeevarajan, Judith and De Dios, Yiri and Cohen, Helen S. and Bloomberg, Jacob J. and Mulavara, Ajitkumar P.},
  journal={PLOS ONE},
  volume={10},
  number={8},
  pages={e0136335},
  year={2015},
  doi={10.1371/journal.pone.0136335}
}

@article{GVSeffect,
  author  = {Fitzpatrick, Richard C. and Day, Brian L.},
  title   = {Probing the human vestibular system with galvanic stimulation},
  journal = {Journal of Applied Physiology},
  year    = {2004},
  volume  = {96(6)},
  pages   = {2301--2316},
  doi     = {10.1152/japplphysiol.00008.2004}
}

@article{GVSwardman2003effects,
  title={Effects of galvanic vestibular stimulation on human posture and perception while standing},
  author={Wardman, Daniel L. and Taylor, Janet L. and Fitzpatrick, Richard C.},
  journal={The Journal of Physiology},
  volume={551},
  number={3},
  pages={1033--1042},
  year={2003},
  doi={10.1113/jphysiol.2003.045971}
}

@article{GVSgallagher2023quantifying,
  title={Quantifying virtual self-motion sensations induced by galvanic vestibular stimulation},
  author={Gallagher, M. and Romano, F. and Bockisch, C. J. and Ferr{\`e}, E. R. and Bertolini, G.},
  journal={Journal of Vestibular Research},
  volume={33},
  number={1},
  pages={21--30},
  year={2023},
  doi={10.3233/VES-220031}
}

@article{GVSfeedback,
  author  = {Smith, Kieran J. and Datta, Abhishek and
             Burkhart, Cody and Clark, Torin K.},
  title   = {Efficacy of Galvanic Vestibular Stimulation as a Display
             Modality Dissociated from Self-Orientation},
  journal = {Human Factors},
  year    = {2024},
  volume  = {66(3)},
  pages   = {862--871},
  doi     = {10.1177/00187208221119879}
}

@article{Safety,
title = {Low intensity transcranial electric stimulation: Safety, ethical, legal regulatory and application guidelines},
journal = {Clinical Neurophysiology},
volume = {128},
number = {9},
pages = {1774-1809},
year = {2017},
issn = {1388-2457},
doi = {https://doi.org/10.1016/j.clinph.2017.06.001},
author = {A. Antal and I. Alekseichuk and M. Bikson and J. Brockmöller and A.R. Brunoni and R. Chen and L.G. Cohen and G. Dowthwaite and J. Ellrich and A. Flöel and F. Fregni and M.S. George and R. Hamilton and J. Haueisen and C.S. Herrmann and F.C. Hummel and J.P. Lefaucheur and D. Liebetanz and C.K. Loo and C.D. McCaig and C. Miniussi and P.C. Miranda and V. Moliadze and M.A. Nitsche and R. Nowak and F. Padberg and A. Pascual-Leone and W. Poppendieck and A. Priori and S. Rossi and P.M. Rossini and J. Rothwell and M.A. Rueger and G. Ruffini and K. Schellhorn and H.R. Siebner and Y. Ugawa and A. Wexler and U. Ziemann and M. Hallett and W. Paulus}
}

@article{Base2025,
  author  = {Marchand, Sarah and Langlade, Alba and Legois, Quentin and S{\'e}verac Cauquil, Alexandra},
  title   = {A wide-ranging review of galvanic vestibular stimulation: from its genesis to basic science and clinical applications},
  journal = {Experimental Brain Research},
  year    = {2025},
  volume  = {243},
  number  = {5},
  pages   = {131},
  doi     = {10.1007/s00221-025-07079-8},
  pmid    = {40289049},
  pmcid   = {PMC12034599}
}

@article{Kwan2019,
  author  = {Kwan, Annie and Forbes, Patrick A. and Mitchell, Diana E. and Blouin, Jean-S{\'e}bastien and Cullen, Kathleen E.},
  title   = {Neural substrates, dynamics and thresholds of galvanic vestibular stimulation in the behaving primate},
  journal = {Nature Communications},
  year    = {2019},
  volume  = {10},
  number  = {1},
  pages   = {1904},
  doi     = {10.1038/s41467-019-09738-1}
}

@article{tdcs,
title = {Electrified minds: Transcranial direct current stimulation (tDCS) and Galvanic Vestibular Stimulation (GVS) as methods of non-invasive brain stimulation in neuropsychology—A review of current data and future implications},
journal = {Neuropsychologia},
volume = {48},
number = {10},
pages = {2789-2810},
year = {2010},
issn = {0028-3932},
doi = {https://doi.org/10.1016/j.neuropsychologia.2010.06.002},
author = {Kathrin S. Utz and Violeta Dimova and Karin Oppenländer and Georg Kerkhoff}
}

@article{ramp2018,
  author  = {Khoshnam, Mahta and H{\"a}ner, Daniela M. C. and Kuatsjah, Eunice and Zhang, Xin and Menon, Carlo},
  title   = {Effects of Galvanic Vestibular Stimulation on Upper and Lower Extremities Motor Symptoms in Parkinson's Disease},
  journal = {Frontiers in Neuroscience},
  year    = {2018},
  volume  = {12},
  pages   = {633},
  doi     = {10.3389/fnins.2018.00633}
}

@article{ramp2006,
  author  = {Lepecq, Jean-Claude and De Waele, Catherine and Mertz-Josse, Sophie and Teyss{\`e}dre, Claudine and Tran Ba Huy, Pham and Baudonni{\`e}re, Pierre-Marie and Vidal, Pierre-Paul},
  title   = {Galvanic Vestibular Stimulation Modifies Vection Paths in Healthy Subjects},
  journal = {Journal of Neurophysiology},
  year    = {2006},
  volume  = {95},
  number  = {5},
  pages   = {3199--3207},
  doi     = {10.1152/jn.00478.2005}
}

@article{ramp2019,
  author  = {Nooristani, Mujda and Maheu, Maxime and Houde, Marie-Soleil and Bacon, Benoit-Antoine and Champoux, Fran{\c{c}}ois},
  title   = {Questioning the Lasting Effect of Galvanic Vestibular Stimulation on Postural Control},
  journal = {PLOS ONE},
  year    = {2019},
  volume  = {14},
  number  = {11},
  pages   = {e0224619},
  doi     = {10.1371/journal.pone.0224619}
}

@inbook{xiaoyu,
author = {Zhao, Xiaoyu and Li, Jingjing and Fushimi, Tatsuki and Ochiai, Yoichi},
title = {Exploring Daily Applications of Wearable Galvanic Vestibular Stimulation via a Self-Customizable Toolkit},
year = {2026},
isbn = {9798400723513},
publisher = {Association for Computing Machinery},
address = {New York, NY, USA},
url = {https://doi.org/10.1145/3795011.3797373},
booktitle = {Proceedings of the Augmented Humans International Conference 2026},
pages = {960–963},
numpages = {4}
}

@inproceedings{TimeWaver,
author = {Narasimhan, Sai Shankar and Agarwal, Shubhankar and Akcin, Oguzhan and Sanghavi, Sujay and Chinchali, Sandeep},
title = {Time Weaver: a conditional time series generation model},
year = {2024},
publisher = {JMLR.org},
booktitle = {Proceedings of the 41st International Conference on Machine Learning},
articleno = {1514},
numpages = {28},
location = {Vienna, Austria},
series = {ICML'24}
}

@article{Zhang2024VAEAugmentation,
  author  = {Zhang, Chunfeng and Qin, Hao and Zhang, Yongjun and Jiang, Chongying and Zhang, Di and Deng, Wenyang},
  title   = {Augmenting Time Series Data: An Interpretable Approach with Metric Learning and Variational Autoencoders},
  journal = {International Journal of Electrical Power \& Energy Systems},
  year    = {2024},
  volume  = {161},
  pages   = {110190},
  doi     = {10.1016/j.ijepes.2024.110190}
}

@inproceedings{retrivl,
author = {Liu, Jingwei and Yang, Ling and Li, Hongyan and Hong, Shenda},
title = {Retrieval-augmented diffusion models for time series forecasting},
year = {2024},
isbn = {9798331314385},
publisher = {Curran Associates Inc.},
address = {Red Hook, NY, USA},
booktitle = {Proceedings of the 38th International Conference on Neural Information Processing Systems},
articleno = {91},
numpages = {21},
location = {Vancouver, BC, Canada},
series = {NIPS '24}
}

@inproceedings{SEDiff,
  author    = {Wang, Xiaoda and Han, Kaiqiao and Xu, Yuhao and Luo, Xiao and Sun, Yizhou and Wang, Wei and Yang, Carl},
  title     = {{SE-Diff}: Simulator and Experience Enhanced Diffusion Model for Comprehensive {ECG} Generation},
  booktitle = {The Fourteenth International Conference on Learning Representations},
  year      = {2026},
  note      = {ICLR 2026 Poster},
  url       = {https://openreview.net/forum?id=95ZV35sBDm}
}

@article{Keywan2018,
  author  = {Keywan, Aram and Wuehr, Max and Pradhan, Cauchy and Jahn, Klaus},
  title   = {Noisy Galvanic Stimulation Improves Roll-Tilt Vestibular Perception in Healthy Subjects},
  journal = {Frontiers in Neurology},
  year    = {2018},
  volume  = {9},
  pages   = {83},
  doi     = {10.3389/fneur.2018.00083}
}

@article{Fujimoto2016,
  author  = {Fujimoto, Chisato and Yamamoto, Yoshiharu and Kamogashira, Teru and Kinoshita, Makoto and Egami, Naoya and Uemura, Yukari and Togo, Fumiharu and Yamasoba, Tatsuya and Iwasaki, Shinichi},
  title   = {Noisy Galvanic Vestibular Stimulation Induces a Sustained Improvement in Body Balance in Elderly Adults},
  journal = {Scientific Reports},
  year    = {2016},
  volume  = {6},
  pages   = {37575},
  doi     = {10.1038/srep37575}
}

@article{Aoyama2015,
  author  = {Aoyama, Kazuma and Iizuka, Hiroyuki and Ando, Hideyuki and Maeda, Taro},
  title   = {Four-pole Galvanic Vestibular Stimulation Causes Body Sway About Three Axes},
  journal = {Scientific Reports},
  year    = {2015},
  volume  = {5},
  pages   = {10168},
  doi     = {10.1038/srep10168}
}

@article{Allred2024,
  author  = {Allred, Aaron R. and Austin, Caroline R. and Klausing, Lanna and Boggess, Nicholas and Clark, Torin K.},
  title   = {Human Perception of Self-Motion and Orientation During Galvanic Vestibular Stimulation and Physical Motion},
  journal = {PLOS Computational Biology},
  year    = {2024},
  volume  = {20},
  number  = {11},
  pages   = {e1012601},
  doi     = {10.1371/journal.pcbi.1012601}
}

@article{Wang2022E5,
  title   = {Text Embeddings by Weakly-Supervised Contrastive Pre-training},
  author  = {Wang, Liang and Yang, Nan and Huang, Xiaolong and Jiao, Binxing and Yang, Linjun and Jiang, Daxin and Majumder, Rangan and Wei, Furu},
  journal = {arXiv preprint arXiv:2212.03533},
  year    = {2022}
}

@inproceedings{MacQueen1967,
  author    = {MacQueen, James B.},
  title     = {Some Methods for Classification and Analysis of Multivariate Observations},
  booktitle = {Proceedings of the Fifth Berkeley Symposium on Mathematical Statistics and Probability},
  volume    = {1},
  pages     = {281--297},
  year      = {1967},
  publisher = {University of California Press}
}

@article{Ratcliff2018,
  author  = {Ratcliff, Roger and Voskuilen, Chelsea and Teodorescu, Andrei},
  title   = {Modeling 2-Alternative Forced-Choice Tasks: Accounting for Both Magnitude and Difference Effects},
  journal = {Cognitive Psychology},
  year    = {2018},
  volume  = {103},
  pages   = {1--22},
  doi     = {10.1016/j.cogpsych.2018.02.002}
}

@inproceedings{kingma2014auto,
  title={Auto-Encoding Variational Bayes},
  author={Kingma, Diederik P. and Welling, Max},
  booktitle={Proceedings of the 2nd International Conference on Learning Representations (ICLR)},
  year={2014}
}

@article{AEbase,
author = {G. E. Hinton  and R. R. Salakhutdinov },
title = {Reducing the Dimensionality of Data with Neural Networks},
journal = {Science},
volume = {313},
number = {5786},
pages = {504-507},
year = {2006},
doi = {10.1126/science.1127647}}

@article{zimage2025,
  title={Z-Image: An Efficient Image Generation Foundation Model with Single-Stream Diffusion Transformer},
  author={{Z-Image Team} and Cai, Huanqia and Cao, Sihan and Du, Ruoyi and Gao, Peng and Hoi, Steven and Huang, Shijie and Hou, Zhaohui and Jiang, Dengyang and Jin, Xin and Li, Liangchen and Li, Zhen and Li, Zhong-Yu and Liu, David and Liu, Dongyang and Shi, Junhan and Wu, Qilong and Yu, Feng and Zhang, Chi and Zhang, Shifeng and Zhou, Shilin},
  journal={arXiv preprint arXiv:2511.22699},
  year={2025}
}

\end{document}